# Human Symmetry Uncertainty Detected by a Self-Organizing Neural Network Map

**Birgitta Dresp-Langley[1]*, John M. Wandeto[2]**

[1] ICube Lab UMR 7357 CNRS, Strasbourg University, FRANCE; birgitta.dresp@unistra.fr
[2] Department of Information Technology, Dedan Kimathi University of Technology, KENYA; john.wandeto@dkut.ac.ke
* Correspondence: birgitta.dresp@unistra.fr

**Abstract:** Symmetry in biological and physical systems is a product of self-organization driven by evolutionary processes, or mechanical systems under constraints. Symmetry-based feature extraction or representation by neural networks may unravel the most informative contents in large image databases. Despite significant achievements of artificial intelligence in recognition and classification of regular patterns, the problem of uncertainty remains a major challenge in ambiguous data. In this study, we present an artificial neural network that detects symmetry uncertainty states in human observers. To this end, we exploit a neural network metric in the output of a biologically inspired Self-Organizing Map, the Quantization Error (SOM-QE). Shape pairs with perfect geometric mirror symmetry but a non-homogenous appearance, caused by local variations in hue, saturation, or lightness within and/or across the shapes in a given pair produce, as shown here, longer choice RT for 'yes' responses relative to symmetry. These data are consistently mirrored by the variations in the SOM-QE from unsupervised neural network analysis of the same stimulus images. The neural network metric is thus capable of detecting and scaling human symmetry uncertainty in response to patterns. Such capacity is tightly linked to the metric's proven selectivity to local contrast and color variations in large and highly complex image data.

## 1. Introduction

Symmetry in biological and physical systems is a product of self-organization [1] driven by evolutionary processes and/or mechanical systems under constraints. It conveys a basic feature to living objects, from molecules to animal bodies, or to physical forces acting in synergy to create symmetrical structures [1-6]. In pattern formation, perfect symmetry is a regularity within a pattern the two halves of which are mirror images of each other. In information theory and in particular human information processing [7-11], symmetry is considered an important carrier of information, detected universally by humans from an early age on [12,13]. Human symmetry detection [14, 15] in patterns or shapes involves visual and cognitive processes from lower to higher levels of functional organization [16-24]. Vertical mirror symmetry is a particularly salient form of visual symmetry [23-25], processed at early stages in human vision and producing greater or lesser detection reliability [23] depending on local features of the stimulus display with greater or lesser stimulus certainty. Shape symmetry is a visual property that attracts attention [18] and determines perceived volume [19-22] and perceptual sa-lience [26] of objects represented in the two-dimensional image plane. Aesthetic judgment and choice preference [27,28] are influenced by symmetry, justifying biologically inspired models of symmetry perception in humans [29] under the light of the fact that symmetry is detected not only by primates but also by other species, such as insects, for example [30].

Symmetry may be exploited in pattern detection and classification by neural networks, which may have to learn multiple copies of one and the same object representation displayed in different orientations. Encoding symmetry as a shape prior in the network can, therefore, help avoid redundant computation where the network has to learn to detect the same pattern or shape in multiple orientations [31]. Symmetry-based feature extraction and/or representation [32] by neural networks using deep learning can, for example, help discover the most informative representations in large image databases with only a minor amount of preprocessing [33]. However, despite significant achievements of artificial intelligence in recognition and classification of well-reproducible patterns, the problem of uncertainty still requires additional attention, especially in ambiguous data. An artificial neural network that detects uncertainty states, where a human observer doubts about an image interpretation has been described previously for the case of MEG image data with significant ambiguity [34]. Here in this study, we present an artificial neural network that detects uncertainty states in human observers about shape symmetry. To this end, we exploit a neural network metric in the output of a biologically inspired Self-Organizing Map (SOM). The Quantization Error (SOM-QE) [35-44] is a measure of output variance and quantifies the difference between neural network states at different stages of unsupervised image learning. In our previous studies, we demonstrated functional properties of the SOM-QE such as a sensitivity to the spatial extent, intensity, and sign or color of local image contrasts in unsupervised image classification by the neural network. The metric reliably detects the finest, clinically or functionally relevant variations in local image contrast contents [36-44], often invisible to the human eye [38,39,42-44]. Here it will be shown that the SOM-QE as a neural network state metric reliably captures, or correlates with, varying levels of human uncertainty in the detection of symmetry of shape pairs with varying local color information contents.

Previous work using human two-alternative forced choice decision had shown that local color variations in two-dimensional pattern displays may significantly influence perceived relative distances [45-48]. In this study here, we use shapes that display perfect geometrical vertical mirror symmetry, as described further below here in 2.1., where visual uncertainty about symmetry is introduced by systematic variations in color (hue) and or saturation of local shape elements, leading to lesser or greater amounts of visual information content. The psychophysically determined choice response time, previously shown to directly reflect stimulus uncertainty in terms of a biological or physiological system states [49,50], is exploited as measure of symmetry uncertainty. Response time (RT) measurement is the psychophysical approach to what is referred to as 'mental chronometry' [51,52,53], which uses measurements of the time between a stimulus onset and an immediate behavioral response to the stimulus. Hick's Law [54,55], in particular, established on the basis of experiments in traditional sensory psychophysics, was repeatedly confirmed in studies by others [49-53]. The law directly relates response time to amount of information in the stimulus, and amount of information in the stimulus to uncertainty as a biologically grounded and potentially universal mechanism underlying sensation, perception, attention and, ultimately, decision [49-54]. In its most general form, Hick's Law postulates that, provided the error rate in the given psychophysical task is low, RT increases linearly with the amount of transmitted information (I) in a set of signals or, as here in this study, a visual image configuration. The law extends to a direct relationship between choice RT and stimulus uncertainty (SU), as plotted graphically in Figure 1 for illustration, and predicts that, provided the response error rate is low, RT increases linearly with stimulus uncertainty (SU)

$$RT = a + b\,(SU) \tag{1}$$

There is no evidence that RT in low-level visual decision tasks would be influenced by individual variables such as gender. Inter-individual differences have been found in RT tasks involving higher levels of semantic processing, as in lexical decision and similar higher level cognitive decision tasks [56]. Biological and cognitive ageing is found to correlate with longer RT due to changes in neurotransmitter dynamics in the ageing brain [57]. This age effect on RT is, however, subject to a certain amount of functional plasticity and may be counteracted by training [58]. To avoid potential effects of age in the experiments here, we chose a study population of healthy young male individuals.

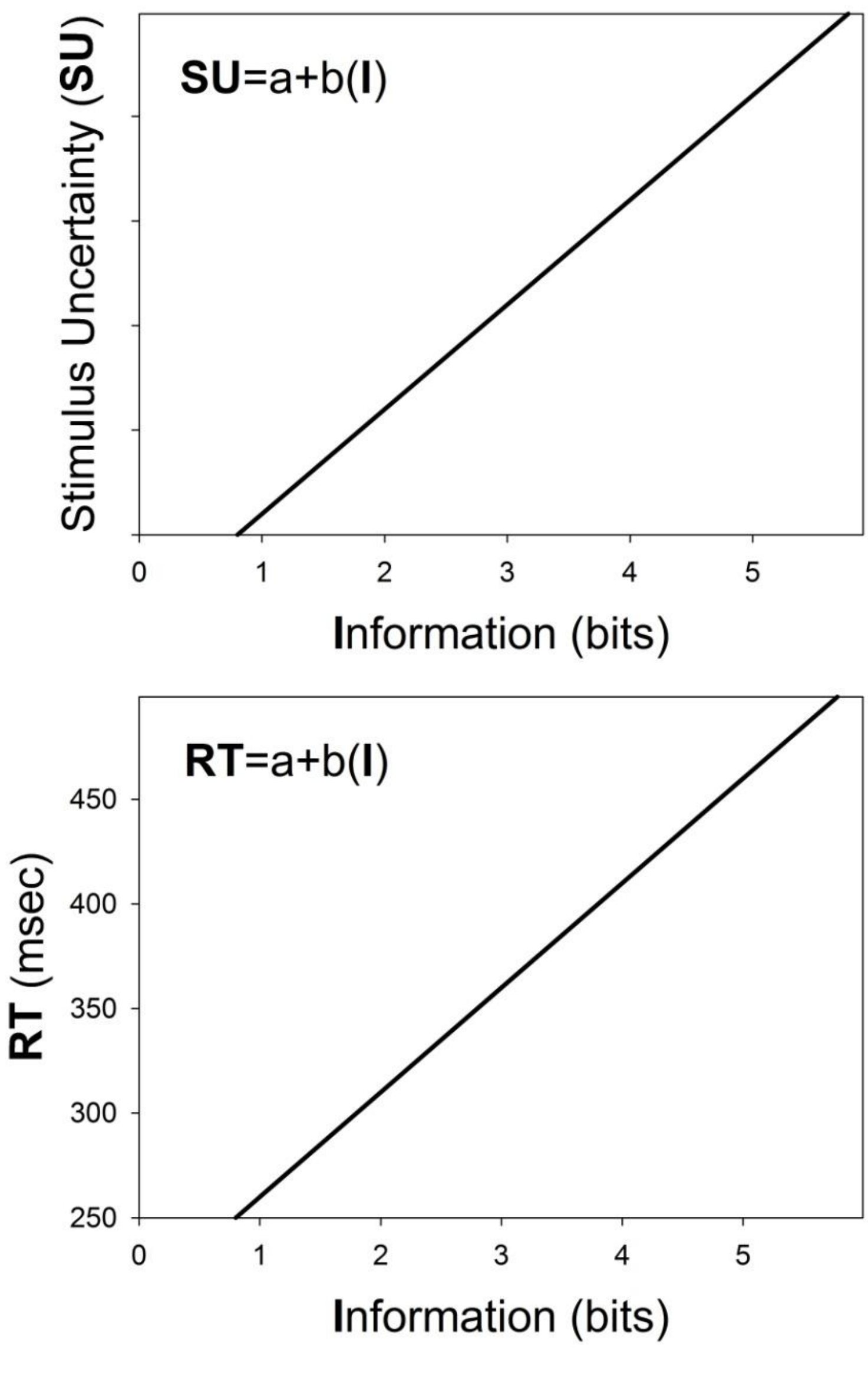

**Figure 1.** Hick's Law [54, 55] postulates that, provided the error rate in the given psychophysical task is low, sensory system uncertainty (SU) increases linearly with the amount of transmitted information (I) in a set (top graph). The law presumes a direct relationship between choice RT and sensory system uncertainty (SU) where RT increases linearly with amount of transmitted information/stimulus uncertainty (bottom graph).

**2. Materials And Methods**

Visual system uncertainty associated with the symmetry of shape pairs was varied experimentally in a series of two-dimensional images showing shape pairs with perfect geometrical (vertical mirror) symmetry but varying amounts of local color information. To quantify human uncertainty in response to the variable amounts of local color information, images were presented in random order on a computer screen to observers who had to decide as quickly as possible whether two shapes in a given images were symmetrical or not (*yes/no* procedure). The psychophysically measured choice response time was computed as measure of uncertainty following the rationale of Hick's Law explained in detail in the introduction here above. The original images where submitted to SOM-QE to test whether the algorithm reliably detects the different levels of human uncertainty reflected by the psychophysical response time variations.

*2.1. Original images*

The original images, fed into the SOM one by one as explained further here below in 2.3., were generated in *Photoshop 12.* They are made available in the Supplementary Materials Section here under 'S1'; copies thereof are shown, for illustration only, in Figure 2. All images are identically scaled. The image size is constant (2720x1446 pixels). All paired shape configurations in the images have perfect geometric mirror symmetry, with the same number of local shape elements on the left and on the right, i.e. a total number of 12 shape elements on each side. All local shape elements are of identical size, the perimeter of each single element, which is the sum of the lengths of the six sides, being constant at 1260 pixels, ensuring constant area size of the local elements. The local color parameters of shape elements in the images were selectively manipulated in Adobe RGB color space to introduce varying levels of visual uncertainty about the mirror symmetry of shape pairs in the different images. The corresponding physical variations in color, hue, saturation, lightness and R-G-B are given here below in Table 1. The medium grey (R=130, G=130, B=130) background is identical in all the images.

*2.2. Experimental display*

The images were displayed for visual presentation on a high resolution computer screen (EIZO COLOR EDGE CG 275W, 2560x1440 pixel resolution) connected to a DELL computer equipped with a high performance graphics card (NVIDIA). Color and luminance calibration of the RGB channels of the monitor was performed using the appropriate Color Navigator self-calibration software, which is delivered with the screen and runs under Windows. As stated here above, visual symmetry uncertainty in the shape pairs was varied by giving the local shape elements variable color appearance in terms of hue, lightness and saturation. We have four variations for each of the single-color shape pair configurations, labeled as 'RED' and 'BLUE', and two variations for the multicolor shape pair configurations, labeled as 'MULTICOL'. As explained in the introduction, we expect that multiple local variations in appearance across shapes of a pair would be likely to produce higher levels of visual symmetry uncertainty in a human observer compared with single local or global variations.

*2.3. Choice Response Time Test*

In the test phase measuring human decision times, the 20 images were displayed in random order in two successively repeated test sessions per human observer. Tests were run on a workstation consisting of a computer screen (EIZO COLOR EDGE CG 275W, 2560x1440 pixel resolution) connected to a DELL computer equipped with a high performance graphics card (NVIDIA). Fifteen healthy young individuals, chosen mainly from a population of undergraduate students, participated in the test phase. All participants had normal or corrected-to-normal visual acuity. In addition, the Ishihara plates [59] were used prior to individual testing to ensure that all of them also had normal color vision.

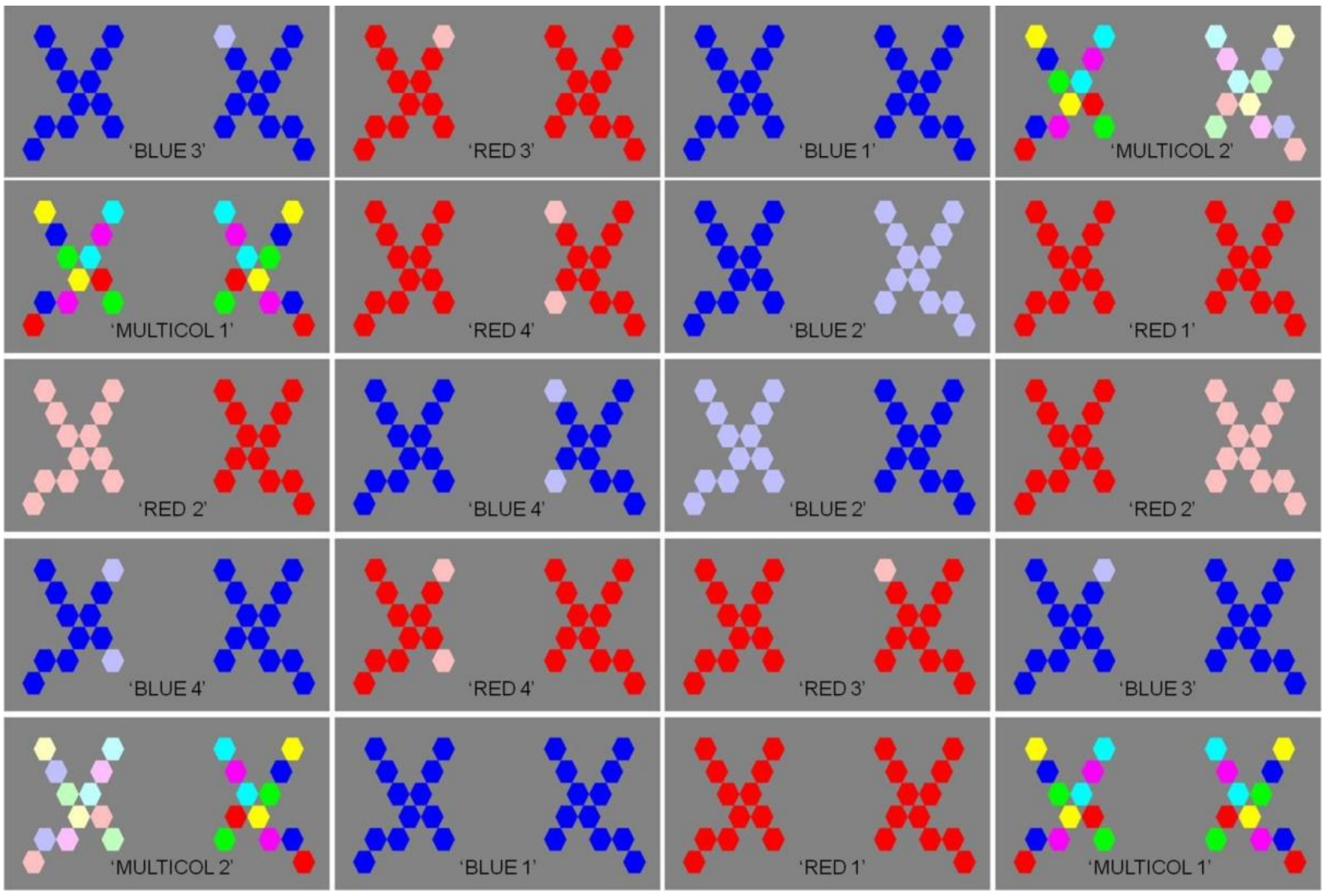

**Figure 2.** Copies of the test images, for illustration. Mirror symmetric shape pairs are displayed on a medium grey background. Visual symmetry uncertainty in the shape pairs was varied by giving shape elements variable amounts of color information resulting in variations in appearance. The condition with the highest amount of locally different color information is MULTICOL2.

**Table 1.** Physical color parameters producing local variations in pattern appearance

| | Color | Hue | Saturation | Lightness | R-G-B |
|---|---|---|---|---|---|
| "Strong" | BLUE | 240 | 100 | 50 | 0-0-255 |
| | RED | 0 | 100 | 50 | 255-0-0 |
| | GREEN | 120 | 100 | 50 | 0-255-0 |
| | MAGENTA | 300 | 100 | 50 | 255-0-255 |
| | YELLOW | 60 | 100 | 50 | |
| "Pale" | BLUE | 180 | 95 | 50 | 10-250-250 |
| | RED | 0 | 100 | 87 | 255-190-190 |
| | GREEN | 120 | 100 | 87 | 190-255-190 |
| | MAGENTA | 300 | 25 | 87 | 255-190-255 |
| | YELLOW | 600 | 65 | 67 | 255-255-190 |

The choice response tests were run in October and November 2019, and in conformity with the Helsinki Declaration for scientific experiments on human individuals. Par-ticipants provided informed consent prior to testing; individuals' identities are not revealed. The test protocol adheres to standards of good procedure stated in the ethics regulations of the CNRS relative to response data collection from healthy humans in non-invasive standard tasks, for which examination of the experimental protocol by a specific ethics committee is not mandatory. Each individual participant was placed in an adjustable chair at a viewing distance of about 80 cm from the computer screen. Individual positions were adjusted in order to ensure that the screen was at eye-level. Tests were run in individual sessions and under *mesopic* viewing conditions. Each participant was adapted to the ambient lighting condition for about five minutes. Instructions given stated that images with two abstract patterns, one on the left and one on the right, would be shown on the screen in two separate sequences. The task communicated to each participant was to: "decide as quickly as possible and as soon as an image appears on the screen whether or not the two patterns in the given image appear to be symmetrical or not". They had to press '1' for '*yes*', or '2' for 'no' on the computer keyboard, their index and middle fingers of their dominant hand ready above the numbers to be able to press a given key without any motor response delay. Each coded response choice was recorded and stored next to the choice response time in a labeled data column of an MS Excel file . The choice response time corresponds to the time between an image onset and the moment a response key is pressed. As soon as a response key was pressed, a current image disappeared from the screen. 900 milliseconds later, the next image was displayed. Image presentations in random order, trial sequencing, and data coding and recording were computer controlled. The program is written in Python for Windows, and freely available for download [60,61].

### *2.4. Neural Network (SOM) Analysis*

The conceptual background and method of neural network analysis follows the same principle and protocol already described in our latest previous work on biological cell imaging data analysis by SOM [37,44]. It is described here again in full detail, for the benefit of the reader. The Self-Organizing Map is an artificial neural network (ANN) with a functional architecture that formally corresponds to the nonlinear, ordered, smooth mapping of high-dimensional input data onto the elements of a regular, low-dimensional array [62]. It is assumed that the set of input variables can be defined as a real vector $x$ of n-dimension. A parametric real vector $m_i$ of n-dimension is associated with each element in the SOM. Vector $m_i$ is a model and the SOM is therefore an array of models. Assuming a general distance measure between $x$ and $m_i$ given by $d(x,m_i)$, the map of an input vector $x$ on the SOM array is defined as the array element $m_c$ that best matches $x$ yielding the smallest $d(x,m_i)$. During the learning process, an input vector $x$ is compared with all the $m_i$ to identify $m_c$. The Euclidean distances $||x\text{-}m_i||$ define $m_c$. Models topographically close in the map up to a certain geometric distance, indicated by $h_{ci}$, will activate each other to learn from their joint input $x$. This results in a local relaxation or smoothing effect on the models in this neighborhood, which in continuous learning leads to global ordering. Learning is represented by

$$m(t+1) = m_i(t) + \alpha(t)h_{ci}(t)[x(t) - m_i(t)] \quad (2)$$

where $t = 1,2,3\ldots$represents an integer, the discrete-time coordinate $h_{ci}(t)$ the neighborhood function - a smoothing kernel defined across the map points which converges towards zero with time - $\alpha(t)$ the learning rate, which also converges towards zero with time and affects the amount of learning in each model. At the end of a *winner-take-all* learning process in the SOM, each image input vector $x$ is matched to its best matching model within the map $m_c$. The difference between $x$ and $m_c$, $||x\text{-}m_c||$, is a measure indicating how close a final SOM value is to the original input value; it is reflected by the quantization error, QE. The average QE of all $x$ (X) within a given image is determined by

$$QE = 1/N \sum_{i=1}^{N} \| X_i - m_{c_i} \| \quad (3)$$

where N is the number of input vectors $x$ in the image. The final weights of the SOM are defined in terms of a three dimensional output vector space representing each R, G, and B channel. Magnitude, as well as the direction of change in any of these from one image to another is reliably reflected by changes in the QE. SOM training here consisted of 1 000 iterations. The SOM was a two-dimensional rectangular map of 4 by 4 nodes, hence capable of creating 16 models, or domains, of observation. The spatial locations, or coordinates, of each model at different locations on the map exhibit specific characteristics, each one different from all the others. When a new input signal is fed into the map, all the models compete; the winner will be the model whose features most closely resemble the features of the input signal. The input signal will be grouped accordingly into one of models. Each model, or domain, is an independent decoder of the same input independently interpreting the information contained in the input, which is represented as a mathematical vector of the same form as that of the model. Therefore, the presence or absence of an active response at a specific map location, rather than the exact input-output signal transformation or magnitude of the response, generates an interpretation of the input. To define initial values for map size, a trial-and-error procedure was implemented. Map sizes larger than 4 by 4 produced observations where some models ended up empty, meaning that these models did not match input the end of the training. As a consequence, 16 models were sufficient to represent all local data in the image. Neighborhood distance and learning rate were assigned initial values of 1.2 and 0.2 respectively. These values were obtained through the trial-and-error procedure, after testing for the quality of the "first guess", directly determined by the value of the resulting quantization error. The smaller the latter, the better the "first guess". It is worthwhile pointing out that the models were initialized by randomly picking training image vectors. This allows the SOM to work on the original data with no prior assumptions about levels of organization within the data. This, however, requires to start with a wider neighborhood function, and a bigger learning-rate factor than in procedures where initial values for model vectors may be pre-selected [63]. Our approach is economical in terms of computation times, which constitutes one of its major advantages. The 20 images here were fed one by one into a single SOM. The training image for the SOM prior to further input can be any of these. Since each of the 20 input images is fed into SOM at a time, it gets subjected to all the 16 models in SOM. These 16 models compete, and the winner ends up as the best matching model for the input vector. After unsupervised *winner-takes-all* SOM learning, the SOM-QE output is written into a data file. Further steps generate output plots of SOM-QE, where each output value is associated with the corresponding input image. The output data are then plotted in increasing/decreasing orders of SOM-QE magnitude as a function of the corresponding image variations (automatic image classification). The computation time of SOM analysis of each of the 20 images was about two seconds per image. The code used for implementing the SOM-QE is available online at:
*https://www.researchgate.net/publication/330500541_Self-organizing_map-based_quantization_error_from_images*

**3. Results**

With 20 images per individual session, two successive sessions per participant, and 15 participants, a total of 600 choice response time data were recorded. A shape pair corresponding to a single factor level relative to shape appearance and color was presented twice in a session with 20 images to allow for left and right hand side presentation of a given appearance factor level in the shape pairs. The labels of the individual factor level associated with each shape pair are given in Figure 1. With two repeated sessions per participants, we have four individual response time data for each single factor level. All data analyses relative to choice response times were run on the 15 average response times for each factor level from the 15 participants.

Since all shape pairs in all the images were mirror-symmetric, *'no'* responses occurred only very rarely in the experiment (17 of the 600 recorded choice responses signaled *'no'*, which corresponds to less than three percent of the total number of observations), as would be expected. In these rare cases, only the choice response times corresponding to a *'yes'* among the four responses recorded for a given factor level were used for computing the average. In terms of operational factor levels in the Cartesian experimental design plan, we have four levels (1,2,3,4) of a factor termed *'Appearance'* ($A_4$) associated with the colors BLUE and RED, and two levels (1,2) of *'Appearance'* ($A_2$) associated with the multiple color case termed MULTICOL here. The three color conditions, blue, red, and multicolor, describe three operational levels of a second factor termed *'Color'* ($C_3$) herein. In a first step, two separate two-way analyses of variance (ANOVA) were run to test for significant effects of the factors *'Appearance'* and *'Color'*. The first ANOVA compares between four levels of *'Appearance'* (1,2,3,4) in two levels (BLUE,RED) of the *'Color'* factor. The second ANOVA compares between two levels of *'Appearance'* in three levels (BLUE, RED, MULTICOL) of the *'Color'* factor.

### *3.1. Two-way ANOVA on choice response times*

#### 3.1.1. $A_4$ x $C_2$ x 15

This analysis corresponds to a Cartesian analysis plan $A_4 \; x \; C_2 \; x \; 15$, with four levels (1,2,3,4) of the '*Appearance'* factor and two levels (BLUE,RED) of the *'Color'* factor on the 15 individual average response times (RT), yielding a total number (N-1) of 119 degrees of freedom (DF). The results from this analysis are shown here below in the top part of Table 2.

**Table 2.** Results from the two-way analyses of variance with factor specific degrees of freedom (DF), the corresponding F statistics, and their associated probability limits (*p*).

| | Factor | DF | F | *p* |
|---|---|---|---|---|
| **1st 2-way ANOVA** | APPEARANCE | 3 | 68.42 | <.001 |
| | COLOR | 1 | .012 | <.914 NS |
| | INTERACTION | 3 | 5.37 | <.01 |
| **2nd 2-way ANOVA** | APPEARANCE | 1 | 8.20 | <.01 |
| | COLOR | 2 | 123.56 | <.001 |
| | INTERACTION | 2 | .564 | <.57 NS |

The results of this analysis signal a statistically significant effect of the *'Appearance'* factor on the average RT and a statistically significant interaction between the *'Appearance'* and the *'Color'* factor for the cases BLUE and RED. A statistically significant effect 'Color' independent of 'Appearance' is not observed, leading to conclude that either of these two colors produced similar effects on RT relative to shape symmetry when their appearance is modified. This holds with the exception for statistical comparison between BLUE3 and BLUE4, which is the only one that is not significant here, as revealed by the *post-hoc* comparison (Holm-Sidak) between these two factor levels (t(1,1) = .32, p<.75 NS). The effects can be appreciated further by looking at the effect sizes for the different conditions, which are visualized further here in the Figures 3 and 4.

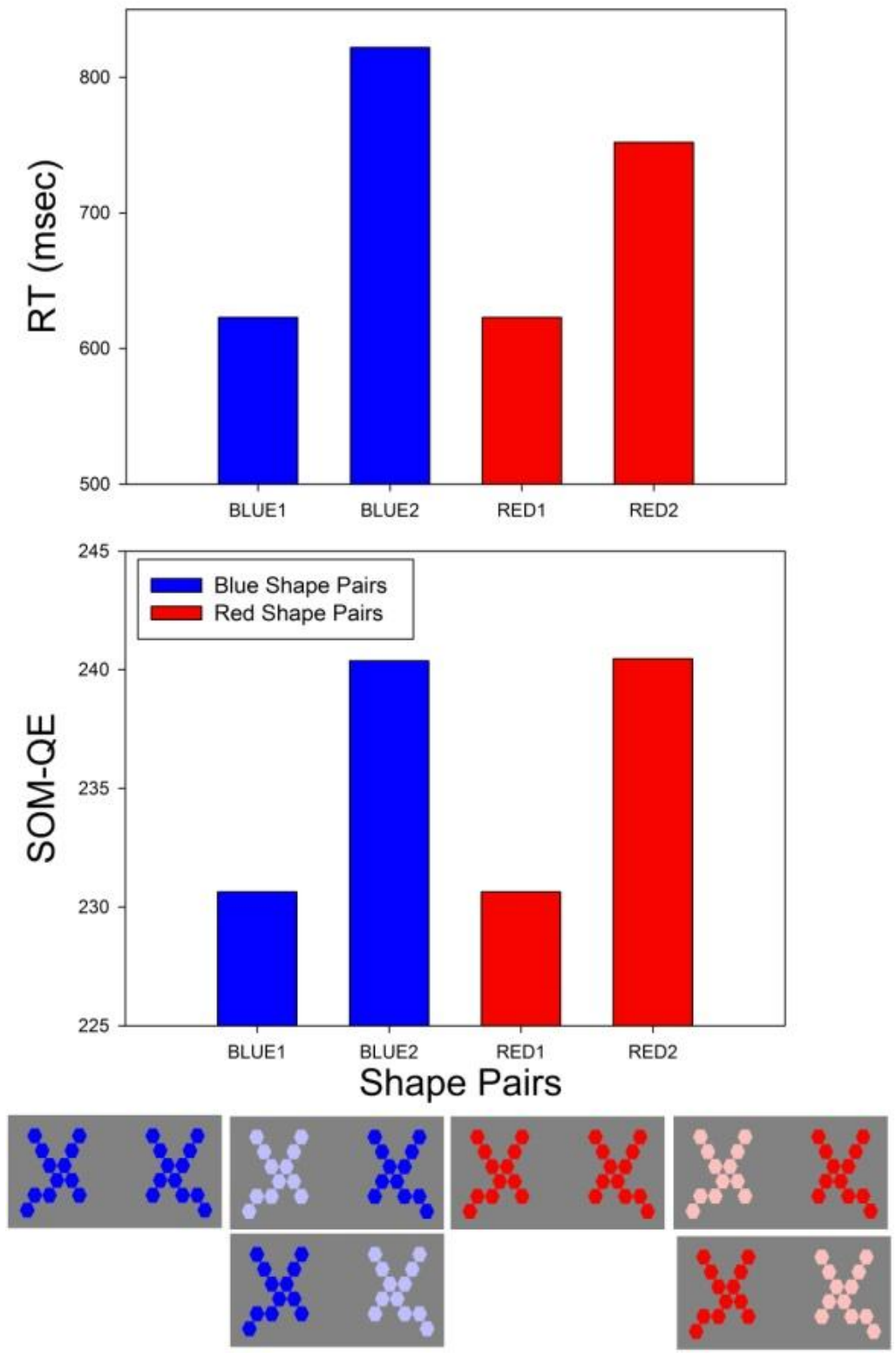

**Figure 3.** Statistically significant differences in average RT (top) for the comparison between BLUE and RED shape pairs with appearance levels 1 and 2. The corresponding SOM-QE values (bottom) from the neural network analysis are plotted in the graph below.

3.1.2. $A_2$ x $C_3$ x 15

This analysis corresponds to a Cartesian analysis plan $A_2 \, x \, C_3 \, x \, 15$, with two levels (1,2) of the *'Appearance'* factor and three levels (BLUE,RED,MULTICOL) of the *'Color'* factor on the 15 individual average response times (RT), yielding a total number (N-1) of 89 degrees of freedom. The results from this analysis are shown here above in the bottom part of Table 2. They signal a statistically significant effect of the *'Appearance'* factor on the average RT and a statistically significant effect of the *'Color'* factor. A statistically significant interaction is not observed here. Statistical *post-hoc* comparisons (*Holm-Sidak*) reveal statistically significant differences between the factor levels MULTICOL and RED ($t(1,1) = 14.44$, $p<.001$) and between the factor levels MULTICOL and BLUE ($t(1,1) = 12.60$, $p<.001$), but not, as could be expected from the previous ANOVA, between the factor levels RED and BLUE ($t(1,1) = 1.84$, $p<.09$ NS). The 'Color' effect here is reflected by the observation that shape pairs with multiple color elements yield significantly longer symmetry related RT compared with shape pairs composed of any of the two single colors here. This effect can be appreciated further by looking at the effect sizes for the different comparisons, which are visualized further below here in Figure 5.

*3.2. RT effect sizes*

The effect sizes, in terms of differences between means, that correspond to significant statistical differences signaled by two-way ANOVA were plotted graphically, and are shown in the top graph in Figure 2 here above, and in the top graphs in Figures 3 and 4 here below along with the corresponding shape pairs that produced the results. The graphs show clearly that shape pairs with non-homogenous appearance, i.e. local variations in hue, saturation, or lightness within and/or across shapes in a given pair, produce longer choice RT for *'yes'* responses relative to shape symmetry.

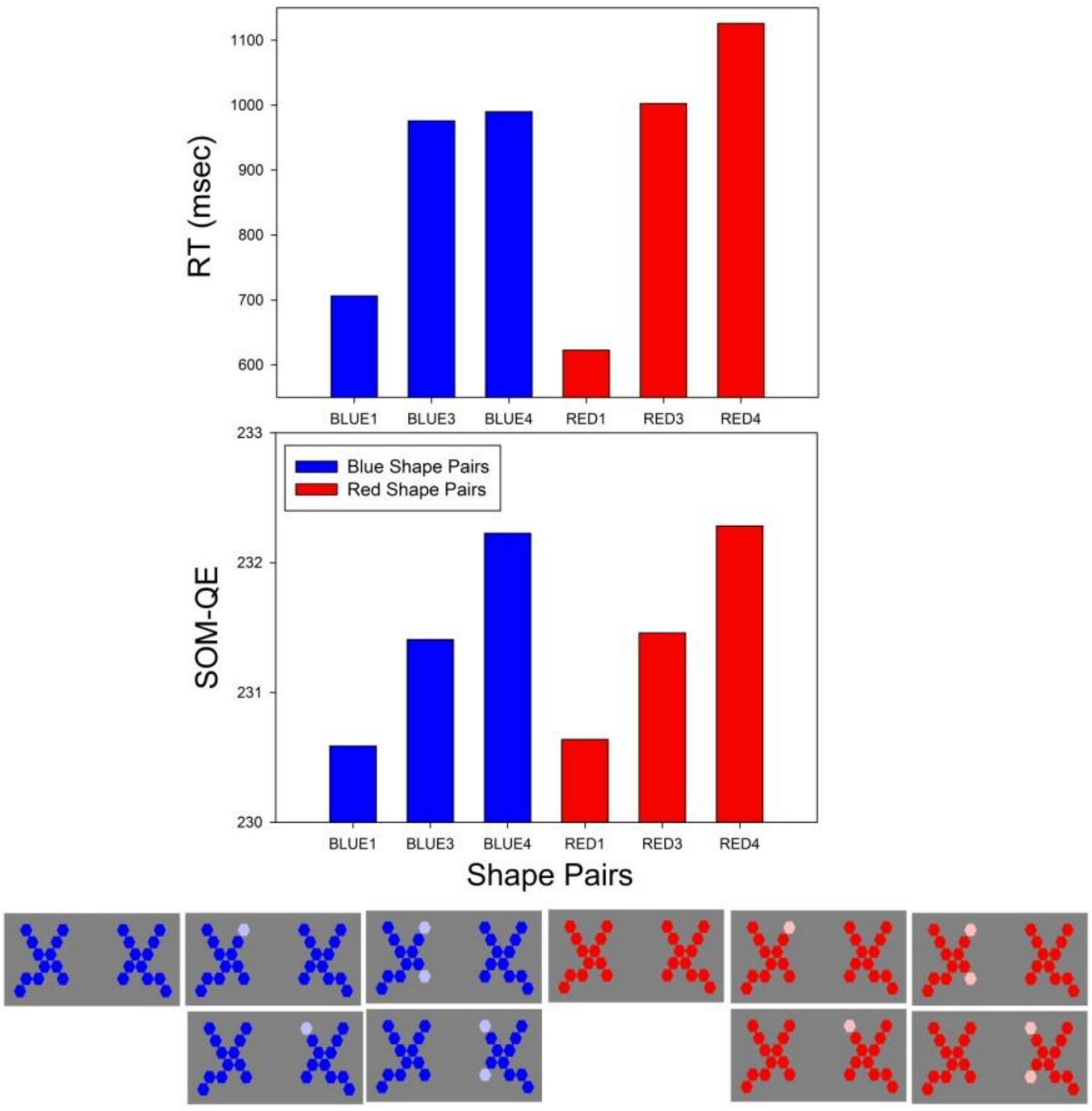

**Figure 4.** Statistically significant differences in average RT (top) for the comparison between BLUE and RED shape pairs with appearance levels 1, 3 and 4. The corresponding SOM-QE values (bottom) from the neural network analysis are plotted in the graph below. The difference in average RT between BLUE3 and BLUE4 is the only one here that is not statistically significant (see paragraph 3.1.1.).

*3.3. SOM-QE effect sizes*

The SOM-QE metrics from the unsupervised neural network analysis of the test images were also plotted graphically and are displayed in the bottom graphs of Figures 2, 3, and 4. The graphs show clearly that the magnitudes of the SOM-QE from the neural network analysis consistently mirror the observed magnitudes of average choice RT for *'yes'* responses relative to shape symmetry produced by shape pairs with varying appearance in terms of local variations in hue, saturation, or lightness within and/or across shapes in a given pair.

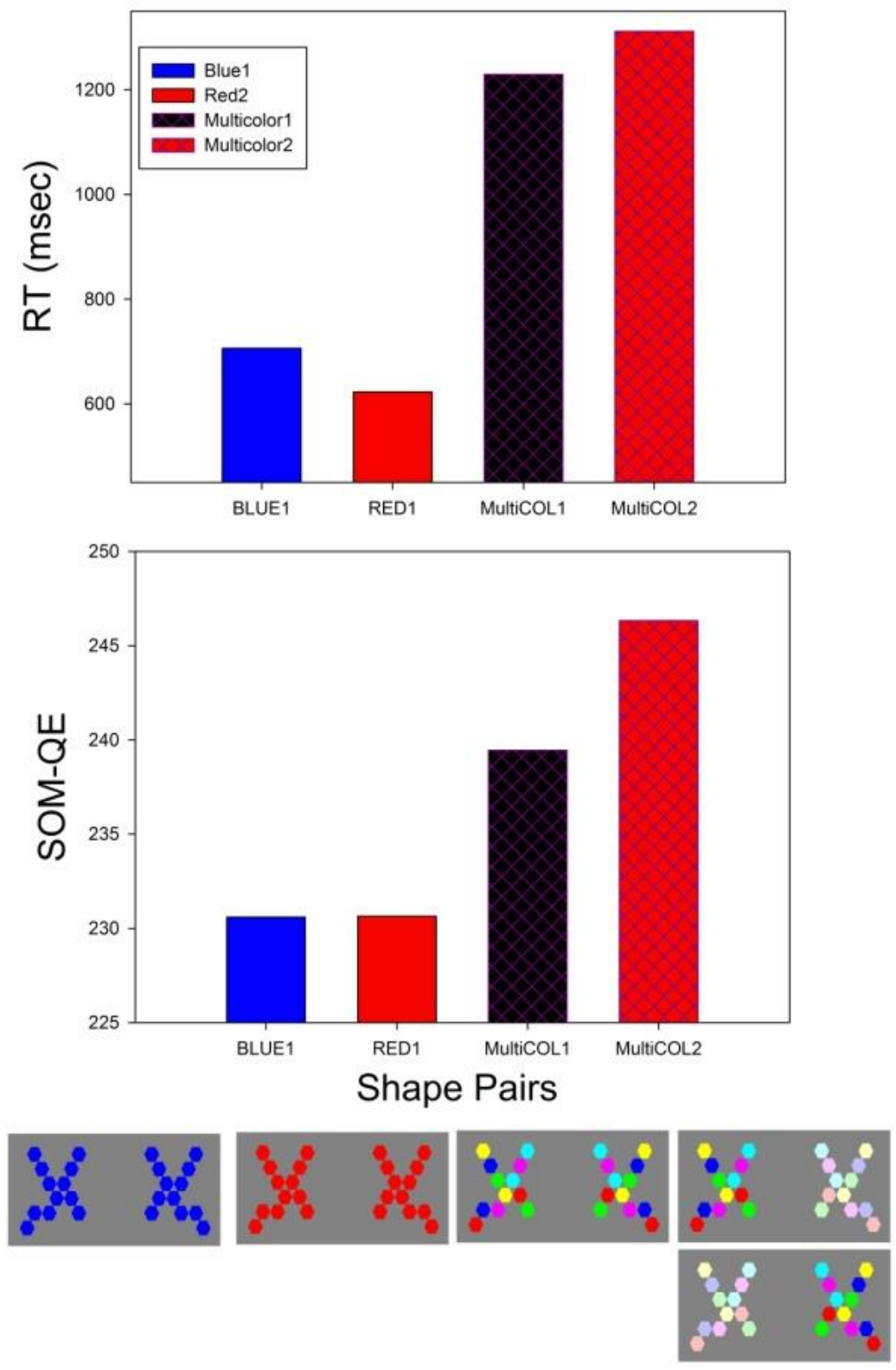

**Figure 5.** Differences in average RT (top) for the comparison between BLUE and RED shape pairs with appearance level 1 and the multicolored MULTICOL shape pairs with appearance levels 1 and 2. The differences between BLUE and RED shape pairs of any appearance level are not statistically significant (see paragraph 3.1.1.). The differences between image conditions BLUE1 or RED1 and MULTICOL1 and between BLUE2 or RED2 and MULTICOL2 are highly significant, as is the difference between MULTICOL1 and MULTICOL2 (see paragraph 3.1.2.). The corresponding SOM-QE values (bottom) from the neural network analysis are plotted in the graph below.

*3.4. Linear regression analyses*

The results from the previous analyses show that the average choice RT for *'yes'* responses relative to shape symmetry, produced by shape pairs with varying appearance in terms of local variations in hue, saturation, or lightness within and/or across shapes in a given pair, produce significant variations consistent with variations in decisional uncertainty about the mirror symmetry of the shapes in a pair. The higher the variability in hue, saturation or lightness of single shape elements, the longer the RT for 'yes" hence the higher the stimulus uncertainty for 'symmetry'. Indeed, the longest choice RT for *'yes'* responses relative to shape symmetry is produced by the shape pairs MULTICOL1 and MULTICOL2. To bring the tight link between variations in RT reflecting different levels of human uncertainty and the variations in the SOM-QE metric from the neural network analyses, we performed a linear regression analysis on the RT data for shape pairs with varying levels of appearance in BLUE, RED and MULTICOL shapes, and a linear regression analysis on the SOM-QE data for exactly the same shape pairs. The results from these analyses are plotted here below in Figure 6. The linear regressions coefficents (R2) are provided in the graph for each analysis. It is shown that RT for *'yes'* responses relative to shape symmetry and the SOM-QE as a function of the same shape variations follow highly similar and signficant linear trends.

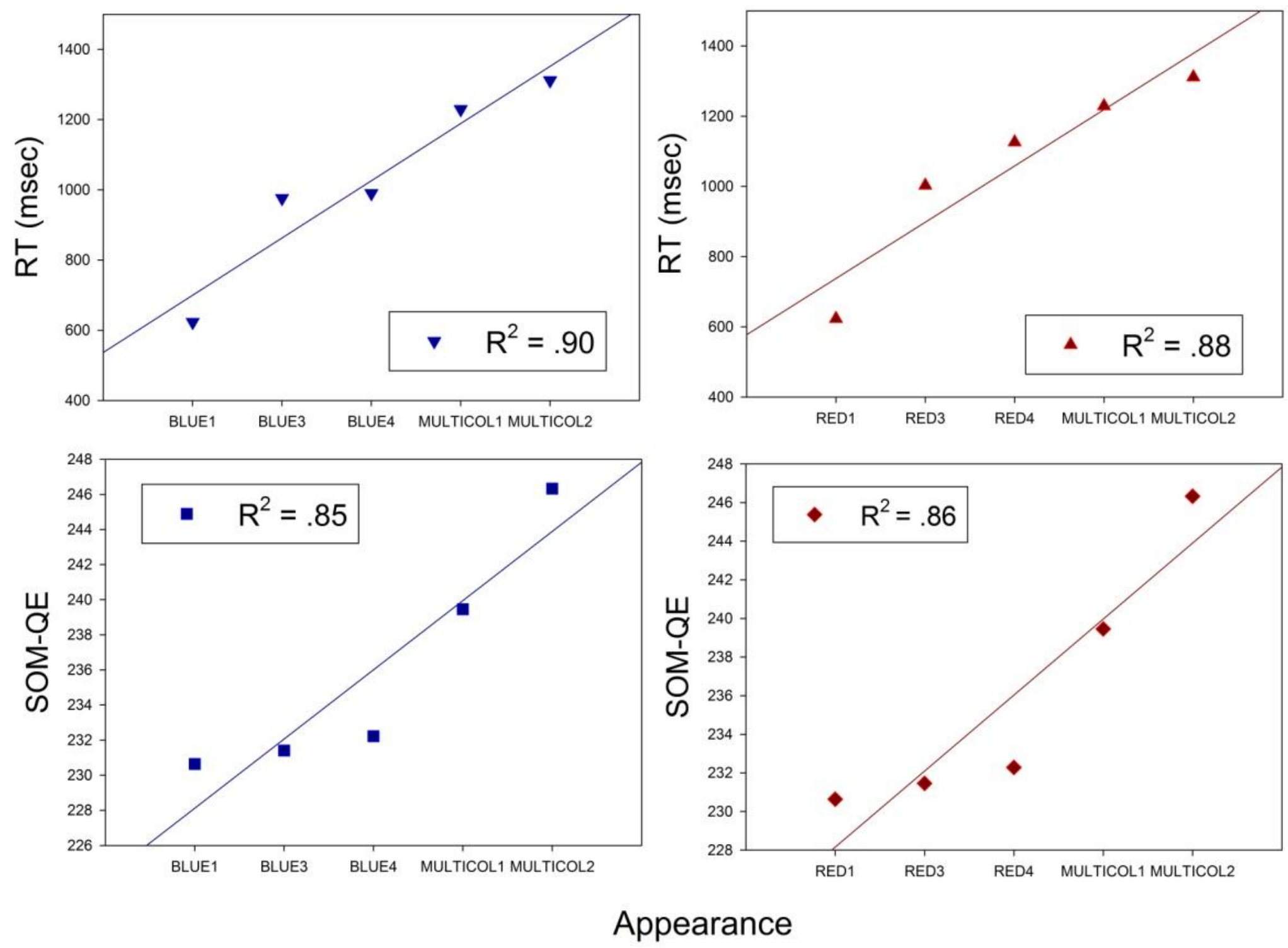

**Figure 6.** The tight link between variations in RT reflecting different levels of human uncertainty and the variations in the SOM-QE metric from the neural network analyses is brought to the fore here under the light of linear regression analysis on the RT data for shape pairs with varying levels of appearance in BLUE, RED and MULTICOL shapes, and linear regression analysis on the SOM-QE data for exactly the same shape pairs.

## 4. Discussion

It is shown that mirror symmetric shape pairs with variable appearance caused by local variations in color information within and/or across shapes of a pair produce longer choice RT for *'yes'* responses relative to the shape symmetry. The variations in average choice RT for *'yes'* are consistent with variations in visual system uncertainty for symmetry, made fully operational here by the variations in amount or diversity of local color information ('appearance') in the display. Consistently, the display with the highest amount of local variations, MULTICOL2, produces the longest choice RT. The higher the the amount of variable information, the longer the RT for 'yes" engendered by the stimulus uncertainty. Although color *per se* may be deemed irrelevant to the given task, color singletons, like the ones manipulated in this experiment here to introduce variations in uncertainty, can act as powerful distractors in a variety of psychophysical choice tasks [64-66]. This is accounted for by models where the attentional weight of any object may be conceived as the product of bottom-up (visual) and top-down (expectational) components [64]. The higher the number of different color singletons in a given paired configuration, the longer the choice response time for "symmetry", in consistency with the most general variant of Hick's Law [54]. As also shown, the variations in RT are consistently mirrored by the variations in the SOM-QE from the unsupervised neural network analysis of the same stimulus images. This provides further data showing that artificial neural networks are capable of detecting human uncertainty in perceptual judgment tasks [34]. The capability of the SOM-QE to capture such uncertainty in human choice responses to the symmetry of shapes with local variations in color parameters is tightly linked to the proven selectivity of this neural network metric to local contrast and color variations in large variety of complex image data [36-44]. Here, the metric is revealed as a measure of both variance in the image input data, and uncertainty in specific human decisions in response to such data. The neural network metric captures the effects of local color contrast on symmetry saliency in cases where pure shape geometry signals perfect mirror symmetry. This unambiguously shows that visual parameters beyond stimulus geometry [67-70] influence what has previously been termed the "symmetry of things in a thing". Such local, non-geometrically determined effects on perceived shape symmetry have potentially important implications for image-guided human precision tasks [71,72], now more and more often assisted by neural network-driven image analysis. From a general functional viewpoint, local-color-based visual system uncertainty for symmetry delaying conscious choice response times in humans is consistent with current theory invoking interactions between low-level visual and higher level cognitive mechanisms in perceptual decisions relative to symmetry [73,74]. The fully conscious detection of symmetry in choice response tasks most likely involves information processing through brain networks with neurons that display large receptive field areas and a massive amount of lateral connectivity [75]. Other machine learning approaches have exploited low-level and global features of natural images or scences to train artificial symmetry detectors on the basis of symmetry-axis ground-truth datasets [76,77], demonstrating the importance of local features such as color in boosting symmetry detection learning by machines compared with gray-scale image data. In human perception, however, the attentional weight of visual information enabling symmetry detection depends not only on the contrast of local features to their local surroundings (saliency), but also on the importance of the features with respect to the given task (relevance). This study here deals with human perceptual system uncertainty when geometrically defined local (feature) and global (shape) symmetry rules are satisfied. Other types of symmetry uncertainty detection, not addressed in this study here, are relevant in image processing. For example, feature selection methods such as Uncertainty Class-Feature Association Map-based selection [78] have proven highly useful preprocessing approaches for eliminating irrelevant and redundant features from complex image data such as DNA microarray data [79,80], where the number of dimensions increase steadily and fast. In such feature selection approaches, which are generally applied to image clusters, the Symmetric Uncertainty Principle [78] is exploited to achieve dimensionality reduction for optimal classification performance. The SOM used in this study here operates on the basis of the assumption that all data in the image (or a cluster thereof) are relevant; the dimensionality of the input to be mapped is predefined and already minimized.

## 5.Conclusions

While symmetry detection by machines and the importance of colored features for such purpose is an increasingly popular field of investigation, questions relating to image properties that may increase system uncertainty for detecting or responding to symmetry, especially when the latter is geometrically perfect, are less often investigated. This work here shows a clear study case where increasing amount of local color information delays the human visual system response to symmetry in consistency with predictions of Hick's Law [54,55]. This system uncertainty is reliably captured by the output metric of a biologically inspired artificial neural network, the SOM, which possesses self-organizing functional properties akin to those of the human sensory system, within the limitations of the study scope set here.